\documentclass[intlimits,twoside,a4paper]{article}

\usepackage{amsmath,amssymb}
\usepackage{graphicx}
\usepackage{psfrag}
\def\FigSize{7.25cm}

\usepackage[eqsecnum]{cmpj2}
\articletype{Regular article}

\title[Potts model with correlated disorder]%
{Stability of the Griffiths phase in the 2D Potts model with correlated disorder}
\author{C. Chatelain\refaddr{label1}}
\addresses{
\addr{label1} Groupe de Physique Statistique,
D\'epartement P2M,
Institut Jean Lamour, CNRS (UMR 7198),
Universit\'e de Lorraine, France}

\begin{document}
\maketitle

\begin{abstract}
A Griffiths phase has recently been observed by Monte Carlo simulations
in the 2D $q$-state Potts model with strongly correlated quenched random couplings.
In particular, the magnetic susceptibility was shown to diverge algebraically
with the lattice size in a broad range of temperatures. However, only relatively
small lattice sizes could be considered so one can wonder whether this Griffiths
phase will not shrink and collapse into a single point, the critical point,
as the lattice size is increased to much larger values. In this paper,
the 2D eight-state Potts model is numerically studied for four different
disorder correlations. It is shown that the Griffiths phase cannot be
explained as a simple spreading of local transition temperatures caused
by disorder fluctuations. As a consequence, the vanishing of the latter
in the thermodynamic limit does not necessarily imply the collapse of
the Griffiths phase into a single point. In contrast, the width of the
Griffiths phase is controlled by the disorder strength. However, for
disorder correlations decaying slower than $1/r$, no cross-over to a more
usual critical behavior could be observed as this strength is tuned to
weaker values.
\pacs 64.60.De, 05.50.+q, 05.70.Jk, 05.10.Ln
\end{abstract}

\section{Introduction}
\label{sec1}
The influence of disorder on phase transitions and critical phenomena has
attracted a considerable interest in the last decades. In the absence of
frustration, it is now well established that a first-order phase transition
is smoothed by the introduction of randomness and can be made continuous
at large enough disorder strength~\cite{ImryWortis}. In 2D, an infinitesimal
disorder is sufficient to remove any discontinuity~\cite{HuiBerker,%
Aizenman,Aizenman2}. When the pure system undergoes already a continuous
phase transition, the Harris criterion predicts that the universality class
of the pure model will be affected by disorder
if the specific heat diverges, i.e. if the critical exponent $\alpha$
is positive~\cite{Harris74}. In this context, the $q$-state Potts model
has been a useful toy model, because it displays a rich phase diagram
involving two lines of respectively first and second-order phase transition.
Along the latter, the universality class depends on the number of states $q$.
On the practical side, efficient Monte Carlo and transfer matrix
algorithms exist for this model and Conformal Invariance can be used in
2D in combination with Renormalisation Group (RG).
 
In comparison, correlated disorder was much less studied. Nevertheless
in some experimental situations, impurities cannot be considered as
uncorrelated. This is in particular the case when they carry an electric
charge or a magnetic moment and are coupled via an electromagnetic interaction.
On the theoretical side, Weinrib and Halperin studied the $\phi^4$ model
with a random mass and showed that a new RG fixed point, distinct from the
random and the pure ones, emerges in the phase diagram when the correlations
of this mass decay algebraically~\cite{Weinrib}.
For a sufficiently slow decay of these disorder correlations,
the new fixed point becomes stable. Denoting $a$ the exponent of the
algebraic decay of disorder correlations, the perturbation is relevant
when $a<d$ if the correlation length exponent $\nu$ of the pure model satisfies
the inequality
%\begin{equation}
%\nu<{2\over a},\quad (a<d).
%\end{equation} 
$\nu<2/a$.
At the new fixed point, correlated disorder is marginally irrelevant,
which implies that that $\nu=2/a$~\cite{Honkonen}. The magnetic exponent
$\eta$ remains small compared to $\epsilon=4-d$. Even though still
controversial, these predictions were confirmed by Monte Carlo simulations
of the 3D Ising model with $a=2$~\cite{IMCorrele,IMCorrele2}.

We recently studied by large-scale Monte Carlo simulations the influence
of correlated couplings on the 2D Potts model~\cite{EPL,PRE}.
Like in the absence of disorder correlation, the first-order phase transition,
occurring for the  pure model when $q>4$, was shown to be smoothed and replaced
by a continuous transition. However, the new universality class was shown
to be $q$-independent, a feature shared by the strong-disorder fixed point
of the $q$-state Potts model with a layered McCoy-Wu-like disorder. This
result is remarkably different from the continuous increase of the
magnetic scaling dimension $x_\sigma$ observed for the Potts model with
an uncorrelated disorder. More intriguing is the fact that the phase diagram
displays a Griffiths phase, as in the McCoy-Wu model, where the magnetic
susceptibility diverges with the lattice size. Interestingly, such a phase
has been predicted by Weinrib and Halperin, but only above the upper
critical dimension $d_c=4$. Finally, the hyperscaling relation ${\gamma\over\nu}
=d-2{\beta\over\nu}$ was observed to be broken in the Griffiths phase,
as a result of large disorder fluctuations.

However, these observations were made for finite systems so one cannot exclude
the possibility that, at much larger lattice sizes, the Griffiths phase
collapse into a single point, the critical point, where the hyperscaling
relation would be restored. Moreover, the estimate of the correlation length
exponent $\nu$ is incompatible with Weinrib-Halperin exact result $\nu=2/a$,
which substantiates the idea that the lattice sizes considered could be too
small and that a cross-over would be observed at much larger lattice sizes.
On the other hand, no significant evolution of the Griffiths phase could be
observed in the range of lattice sizes studied~\cite{PRE}. Moreover, the
conspiracy of two amplitudes that leads to the violation of the hyperscaling
relation is well verified and no sign of deviation at large lattice
sizes is observed.

Since larger lattice sizes are not accessible by Monte Carlo simulations,
we turn our attention in this work to larger exponents $a$ of the disorder
correlations. The fact that Weinrib-Halperin predictions were confirmed by
Monte Carlo simulations of the 3D Ising model with $a=2$ could indicate
that finite-size effects are weaker for larger values of $a$.
In refs~\cite{EPL,PRE}, only small values of $a$ were considered
because disorder configurations were generated by simulating an auxiliary
Ashkin-Teller model on a self-dual line where its critical exponents
are known exactly. The polarisation density was then used to construct the
couplings $J_{ij}$ of the Potts model. Disorder correlations correspond
therefore to the polarisation-polarisation correlations of the auxiliary
Ashkin-Teller model. When moving along the self-dual line, only exponents
in the range $a\in [1/4;3/4]$ can be obtained.

In this work, we present new data for disorder correlation exponents
$a=1/3$ and $2/3$ obtained by using an auxiliary Ashkin-Teller model.
In order to investigate the possible existence of a cross-over towards the
Weinrib-Halperin fixed point, we considered also the values $a\simeq 1.036$
and $a=2$ obtained using the 3D and 4D Ising models as auxiliary models to
generate the disorder configurations. In the first section, details about
the models and the Monte Carlo simulations are given. In the second section,
the behaviour of the magnetic susceptibility $\bar\chi$ is discussed.
As already observed in Ref.~\cite{PRE}, $\bar\chi$ diverge
algebraically with the lattice size in a broad interval of temperatures,
identified as a Griffiths phase, when $a$ is sufficiently small.
A simple explanation of this phenomena is to assume that disorder
fluctuations induce a spreading of local transition temperatures. Because
these fluctuations vanish as $L^{-a/2}$, this would imply that a single
peak would be recovered at large lattice sizes. Moreover, the smaller the
exponent $a$ and the larger the lattice sizes needed to observe a single peak.
%This would explain why the observed effect is stronger when the
%disorder correlation exponent $a$ is small.
In the second section,
numerical evidence is given that disorder fluctuations are not sufficient
to explain the observed Griffiths phase, and therefore, that the latter
phase cannot be expected to collapse as $L^{-a/2}$. In the third section,
the possibility of a cross-over controlled by the amplitude of disorder
correlations is considered. These amplitudes are compared for the
different disorder correlations considered and, then different disorder
strengths are studied. Finally, conclusions follow.

\section{Models and simulation}
\label{sec2}
The classical $q$-state Potts model is the lattice spin model defined by the
Hamiltonian~\cite{Potts,Wu}
   \begin{equation}
     H=-J\sum_{(i,j)} \delta_{\sigma_i,\sigma_j},\hskip 1truecm
     \sigma_i=0,1,\ldots,q-1
     \end{equation}
where the spin $\sigma_i$ takes $q$ possible values and is located on the
$i$-th node of the lattice. The sum extends over all pairs $(i,j)$ of nearest
neighbours on the lattice. In the following, the Potts model will be considered
on the square lattice. As mentioned in the introduction, the phase transition
is continuous for $q\le 4$ and of first-order when $q>4$. We will restrict
ourselves to the case $q=8$, i.e. a point in the regime of first-order
transition. Disorder is now introduced as bond-dependent random exchange
couplings $J_{ij}$. The Hamiltonian becomes
   \begin{equation}
     H=-\sum_{(i,j)} J_{ij}\delta_{\sigma_i,\sigma_j}.
     \end{equation}
The spatial correlations between these couplings is assumed to decay
algebraically with an exponent $a$ at large distance:
   \begin{equation}
     \overline{J_{ij}J_{kl}}-\overline{J_{ij}}\ \!\overline{J_{kl}}
     \sim |\vec r_i-\vec r_k|^{-a}.
     \end{equation}
For convenience, we will restrict ourselves in the following to a binary
coupling distribution, i.e. $J_{ij}=J_1$ or $J_2$. The presence of disorder
correlations does not affect the self-duality condition of the random Potts
model. Imposing $J_1$ and $J_2$ to be equiprobable and self-dual of each other,
the self-dual line is given by the condition~\cite{Kinzel}
   \begin{equation}
     \big(e^{\beta J_1}-1\big)\big(e^{\beta J_2}-1\big)=q.
   \end{equation}

The coupling configurations are generated by independent Monte Carlo simulations
of two auxiliary models: the Ising and Ashkin-Teller models. The former is
defined by the Hamiltonian
   \begin{equation}
     H=-J_{IM}\sum_{(i,j)} \sigma_i\sigma_j,\hskip 1truecm\sigma_i=\pm 1
     \end{equation}
and is equivalent to the $q=2$ Potts model. It is well known that this model
undergoes a second-order phase transition in any dimension $d>1$. We considered
hypercubic lattices of dimension $d=3$ and $d=4$. A few thousand spin
configurations are generated at the critical point, corresponding to $\beta
J_{IM}\simeq 0.221655$ for $d=3$~\cite{Pelissetto} and $\beta J_{IM}\simeq 0.149694$
for $d=4$~\cite{Stauffer97}. For each spin configuration, a two-dimensional
section is cut and random couplings for the 2D Potts model are constructed as
    \begin{equation}
      J_{ij}={J_1+J_2\over 2}+{J_1-J_2\over 2}\sigma_i
      \end{equation}
for each pair $(i,j)$ of nearest neighbours in the 2D section. Note that,
at any site $i$, two couplings, in two different directions, are identical.
By construction, disorder correlation functions $\overline{J_{ij}J_{kl}}
-\overline{J_{ij}}\ \!\overline{J_{kl}}$ decay as the spin-spin correlation
functions of the auxiliary Ising model. Therefore, the decay is algebraic
at large distances with an exponent $a=2\beta/\nu\simeq 1.036$ for the 3D
Ising model~\cite{Pelissetto} and $a=2$ for the 4D Ising model.
Note that, in the second case, the exponent $a$ is equal to the dimension
$d=2$ of the Potts model. Therefore, according to Weinrib and Halperin,
disorder correlations are expected to be irrelevant and the system falls
into the same universality class as the
Potts model with independent random couplings.

The second auxiliary model is the 2D Ashkin-Teller model defined by the
Hamiltonian~\cite{AshkinTeller,Fan}
   \begin{equation}
     H=-\sum_{(i,j)} \big[J_{AT}\sigma_i\sigma_j+J_{AT}\tau_i\tau_j+
       K_{\rm AT}\sigma_i\sigma_j\tau_i\tau_j\big],\hskip 1truecm
       \sigma_i,\tau_i=\pm 1
     \end{equation}
and corresponding to two Ising models coupled by their energy densities.
On the square lattice, the model is self-dual along the line of the phase
diagram given by $e^{-2K_{\rm AT}}=\sinh 2J_{\rm AT}$. Thanks to a mapping onto
the eight-vertex model, the critical exponents are known exactly along this
line. The random couplings for the Potts model are constructed from the
polarisation density as
    \begin{equation}
      J_{ij}={J_1+J_2\over 2}+{J_1-J_2\over 2}\sigma_i\tau_i.
      \end{equation}
The disorder correlations therefore decay as the polarisation-polarisation
correlation functions of the auxiliary Ashkin-Teller model. In this work, we
considered two points on the self-dual line of the Ashkin-Teller model ($y=0.50$
and $y=1.25$ in the language of the eight-vertex model) corresponding to
exponents $a=1/3$ and $a=2/3$.

The above-described spin models were simulated using Monte Carlo cluster
algorithms to reduce the critical slowing-down. For the Ising and Potts models,
the Swendsen-Wang algorithm was employed~\cite{SW}. The Ashkin-Teller was
simulated using a cluster algorithm introduced by Wiseman and
Domany~\cite{Salas,Salas2}.

\section{Griffiths phase and disorder fluctuations}
\label{sec3}
\begin{figure}[ht]
\centering
\psfrag{chi}[Bc][Bc][1][1]{$\overline{\chi}$}
\psfrag{beta}[tc][tc][1][0]{$\beta$}
\psfrag{L=24}[Bc][Bc][1][0]{\tiny $L=24$}
\psfrag{L=32}[Bc][Bc][1][0]{\tiny $L=32$}
\psfrag{L=64}[Bc][Bc][1][0]{\tiny $L=64$}
\psfrag{L=48}[Bc][Bc][1][0]{\tiny $L=48$}
\psfrag{L=96}[Bc][Bc][1][0]{\tiny $L=96$}
\psfrag{L=128}[Bc][Bc][1][0]{\tiny $L=128$}
\includegraphics[width=\FigSize]{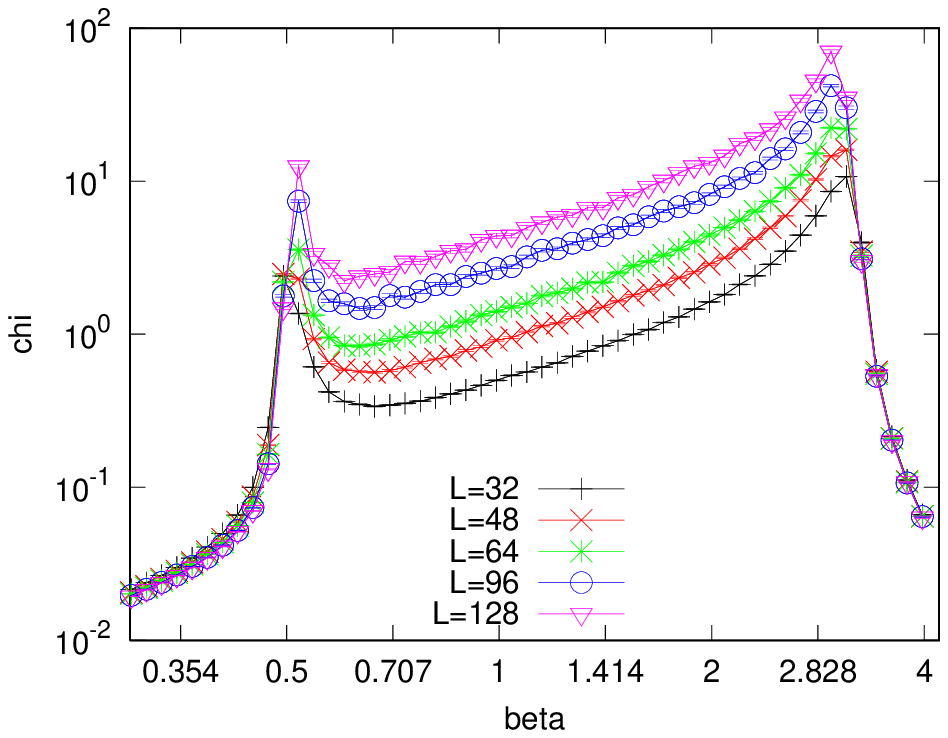}\quad
\includegraphics[width=\FigSize]{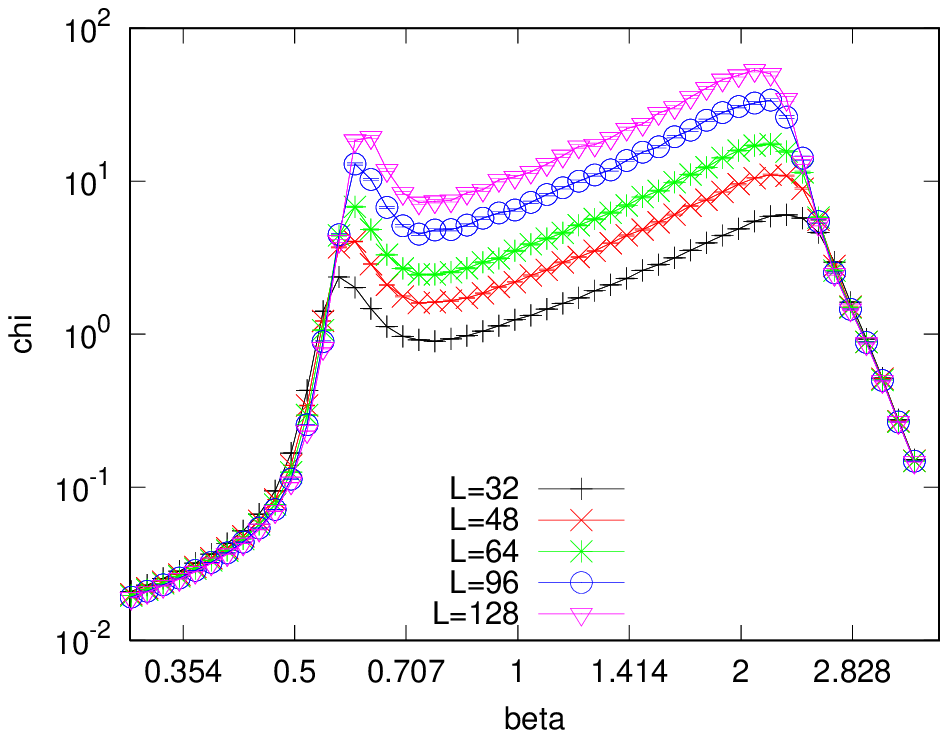}\par
\includegraphics[width=\FigSize]{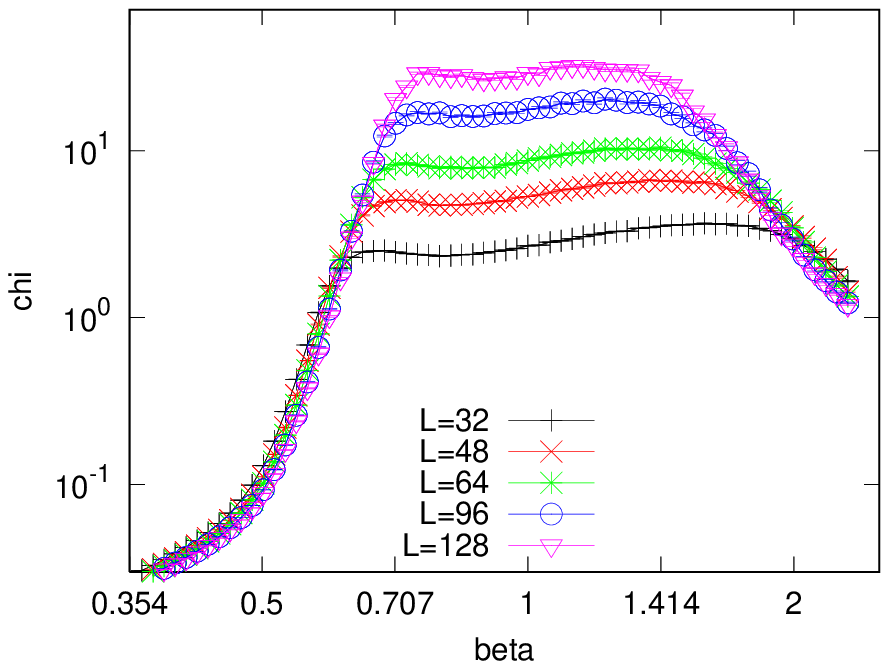}\quad
\includegraphics[width=\FigSize]{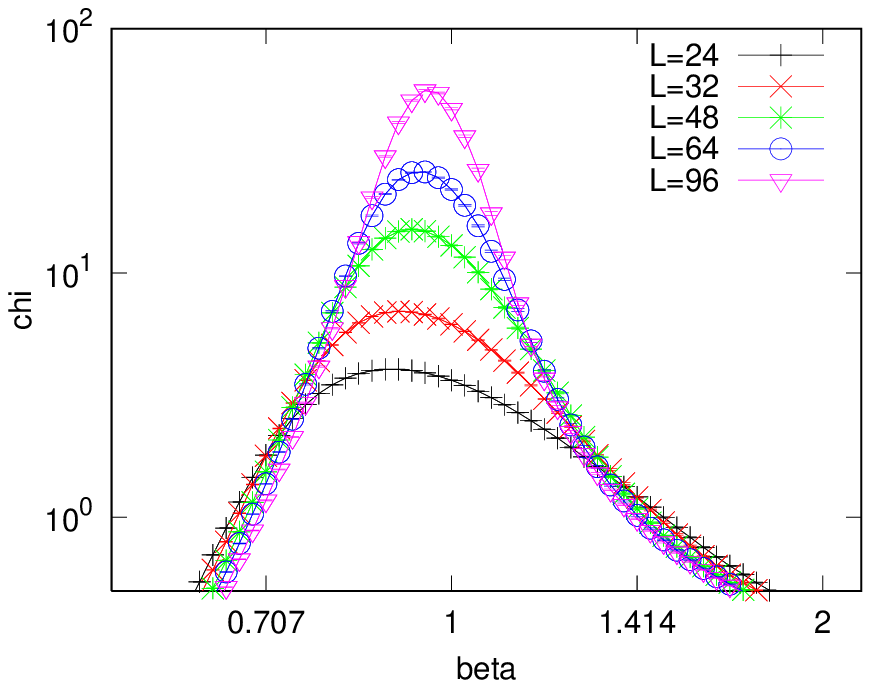}
\caption{Average magnetic susceptibility of the 8-state Potts model
with different disorder correlation exponents ($a=1/3$, $2/3$, $1.036$ and $2$
from left to right and top to bottom) for a disorder strength $r=J_1/J_2=7.5$.
The different curves correspond to different lattice sizes.
Note that the scale of the $y$-axis is logarithmic.}
\label{Fig2}
\end{figure}

The magnetic susceptibility $\chi$ of a finite system undergoing a continuous
phase transition in the thermodynamic limit is expected to display a peak whose
maximum diverges with the lattice size $L$ as $L^{\gamma/\nu}$. The location of
this maximum goes towards the critical temperature $T_c$ in the limit of an
infinite system. A very different situation was observed in the 2D Potts
model with strongly correlated disorder~\cite{PRE}. As can be seen on
figure~\ref{Fig2}, two peaks are present for $a=1/3$ and $2/3$. The data
show an algebraic increase of the average magnetic susceptibility for all
temperatures between these two peaks. For this reason, this region was
conjectured to be a Griffiths phase, similar to the one observed in the
McCoy-Wu model. The absence of any evolution of the location of the two
peaks was reported in the case $a=0.4$. In contrast, figure~\ref{Fig2}
shows a slow evolution in the case $a=2/3$. Since only lattice sizes up
to $L=128$ were studied, the possibility of a collapse of the Griffiths
phase into a single point in the thermodynamic limit cannot be excluded.
Moreover, such a collapse is even more clearly seen on figure~\ref{Fig2}
for $a\simeq 1.036$. Two peaks are still visible but they tend to come closer
when the lattice size is increased. It seems natural in this case to assume
that the two peaks will merge into a single one at larger lattice sizes.
For disorder correlations with a faster decay $a=2$, only one peak is observed
(Fig.~\ref{Fig2}) and its location tends towards the critical value
$\beta_c=1$, expected from self-duality arguments.

% --- MAGNETIZATION --- 
%\begin{figure}
%\centering
%\psfrag{<m>}[Bc][Bc][1][1]{$\overline{\langle m\rangle}$}
%\psfrag{beta}[tc][tc][1][0]{$\beta$}
%\psfrag{L=24}[Bc][Bc][1][0]{\tiny $L=24$}
%\psfrag{L=32}[Bc][Bc][1][0]{\tiny $L=32$}
%\psfrag{L=64}[Bc][Bc][1][0]{\tiny $L=64$}
%\psfrag{L=48}[Bc][Bc][1][0]{\tiny $L=48$}
%\psfrag{L=96}[Bc][Bc][1][0]{\tiny $L=96$}
%\psfrag{L=128}[Bc][Bc][1][0]{\tiny $L=128$}
%\includegraphics[width=\FigSize]{Fig1a.eps}\quad
%\includegraphics[width=\FigSize]{Fig1b.eps}\par
%\includegraphics[width=\FigSize]{Fig1c.eps}\quad
%\includegraphics[width=\FigSize]{Fig1d.eps}
%\caption{Average magnetisation of the 8-state Potts model
%with different disorder correlation exponents ($a=1/3$, $2/3$, $1.036$ and $2$
%from top to bottom and left to right).
%The different curves correspond to different lattice sizes.}\label{Fig1}
%\end{figure}

% --- SUSCEPTIBILITY ---

As mentioned in the introduction, it may be assumed that the width of the
Griffiths phase is due to large disorder fluctuations. It seems indeed natural
to assume that the first peak is caused by the ferromagnetic ordering
of large clusters with a high concentration of weak bonds $J_2$ while
the second one corresponds to clusters of strong bonds $J_1$. Such large
clusters are more probable when disorder correlations decay slowly.
In the following, disorder fluctuations will be compared for the
different values of $a$ considered. To be more specific, consider the
general case of a lattice model with an energy density denoted
$\epsilon_{ij}=\epsilon(\sigma_i,\sigma_j)$ on the edge between the
spins on sites $i$ and $j$. The weak disorder limit of the partition
function can be calculated using the replica trick:
  \begin{equation}
  \overline{\ln{\cal Z}}=\lim_{n\rightarrow 0} {1\over n}\big(
  \overline{{\cal Z}^n}-1\big).
  \end{equation}
Introducing the interaction energy $\epsilon_{ij}^\alpha=\epsilon
(\sigma_i^\alpha,\sigma_j^\alpha)$ between the two spins $\sigma_i^\alpha$
and $\sigma_j^\alpha$ of the $\alpha$-th replica, the partition function
of $n$ replicas reads
  \begin{equation}
  \overline{{\cal Z}^n}=\sum_{\{\sigma_i^\alpha\}}
  \overline{e^{-\beta \sum_{(i,j),\alpha}J_{ij}\epsilon_{ij}^\alpha}}
    \simeq \sum_{\{\sigma_i^\alpha\}}
    e^{-\beta \sum_{(i,j),\alpha}\overline{J_{ij}}\epsilon_{ij}^\alpha
      +{\beta^2\over 2}\sum_{(i,j),(k,l),\atop\alpha,\beta}
      \big(\overline{J_{ij}J_{kl}}-\overline{J_{ij}}\ \!\overline{J_{kl}}\big)
      \epsilon_{ij}^\alpha\epsilon_{kl}^\beta+\ldots}.
    \label{DevPartF}
  \end{equation}
The first contribution of disorder to the partition function involves
the correlations $\overline{J_{ij}J_{kl}}-\overline{J_{ij}}\ \!
\overline{J_{kl}}$, and is obviously a function of $a$. In order to
characterise the disorder strength by a scalar, we considered the sum
of these correlations, which also corresponds to the fluctuations
of the couplings:
   \begin{equation}
     \Delta J_2=\overline{\Big[{1\over N}
       \sum_{(i,j)}(J_{ij}-\bar J)\Big]^2}^{1/2}
     \label{DefSigma}
   \end{equation}
where $N=2L^2$ is the number of bonds of the square lattice. Since the
couplings $J_{ij}$ are constructed from the polarisation density $\sigma_i
\tau_i$ of the auxiliary Ashkin-Teller model (for $a=1/3$ and $2/3$), or
from the magnetisation density $\sigma_i$ of the auxiliary Ising model
(for $a\simeq 1.036$ and $2$), $\Delta J_2$ is related, up to a prefactor
${J_1-J_2\over 2}$, to the fluctuations of the polarisation, or magnetisation,
density. Therefore, $\Delta J_2$ is expected to scale as
   \begin{equation}
     \Delta J_2\sim L^{-a/2}
     \label{ScaleSigma}
   \end{equation}
for both auxiliary models. This result is obtained by expanding the square
in equation~(\ref{DefSigma}) and integrating out the disorder correlations
in the continuum limit. Up to a further factor $L^d$, $(\Delta J_2)^2$ is
also proportional to the electric or magnetic susceptibility of the
Ashkin-Teller and Ising models. The hyperscaling relation for these
auxiliary models leads to $\Delta J_2\sim L^{-\beta/\nu}$ where the
exponents $\beta/\nu$ is equal to $a/2$ by construction of the random
couplings. Equation~(\ref{ScaleSigma}) shows that $\Delta J_2$
behaves as a shift of the critical temperature $|T_c(L)-T_c(\infty)|$
in a finite system. Indeed, one expects $|T_c(L)-T_c(\infty)|
\sim L^{-1/\nu}$ and, at the Weinrib-Halperin fixed point, $\nu=2/a$.

% --- FLUCTUATIONS DU COUPLAGE ---
\begin{figure}[ht]
\centering
\psfrag{s1}[Bc][Bc][1][1]{$\Delta J_1$}
\psfrag{s2}[Bc][Bc][1][1]{$\Delta J_2$}
\psfrag{s4}[Bc][Bc][1][1]{$\Delta J_4$}
\psfrag{L}[tc][tc][1][0]{$L$}
\psfrag{a=1/3}[Bc][Bc][1][0]{\tiny $a=1/3$}
\psfrag{a=2/3}[Bc][Bc][1][0]{\tiny $a=2/3$}
\psfrag{a=1.036}[Bc][Bc][1][0]{\tiny $a\simeq 1.036$}
\psfrag{a=2}[Bc][Bc][1][0]{\tiny $a=2$}
\includegraphics[width=4.85cm]{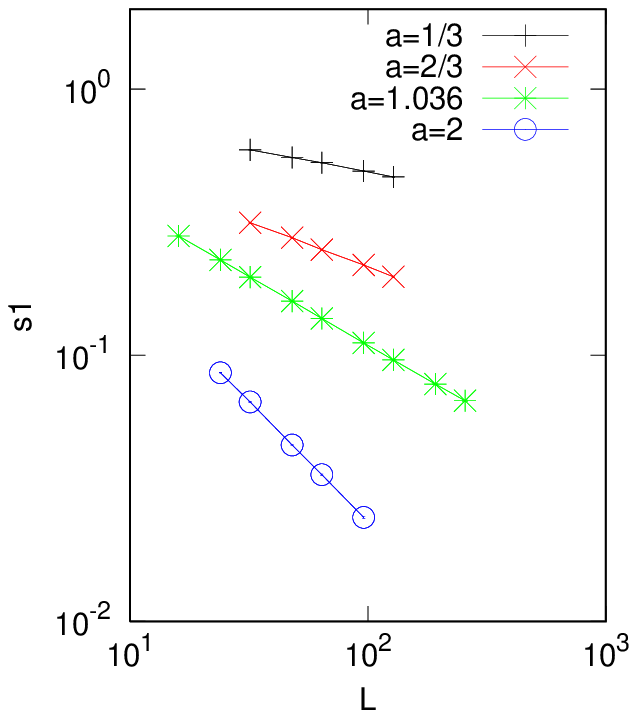}
\includegraphics[width=4.85cm]{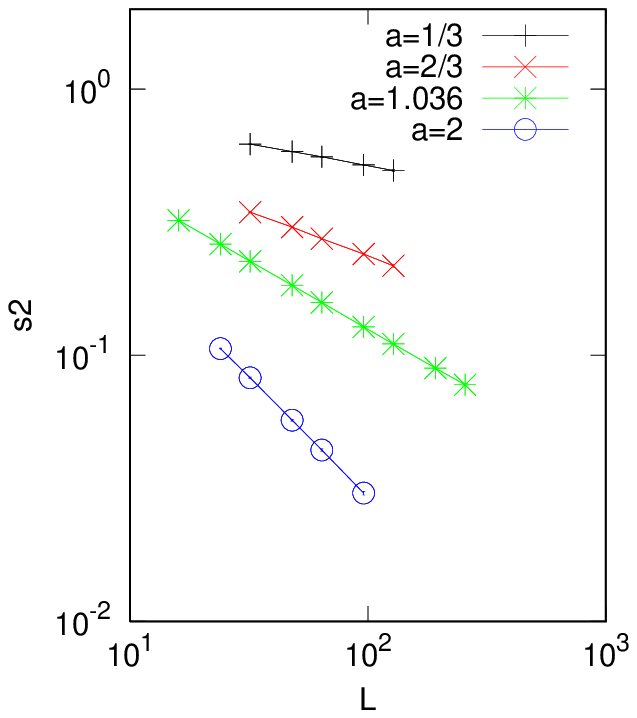}
\includegraphics[width=4.85cm]{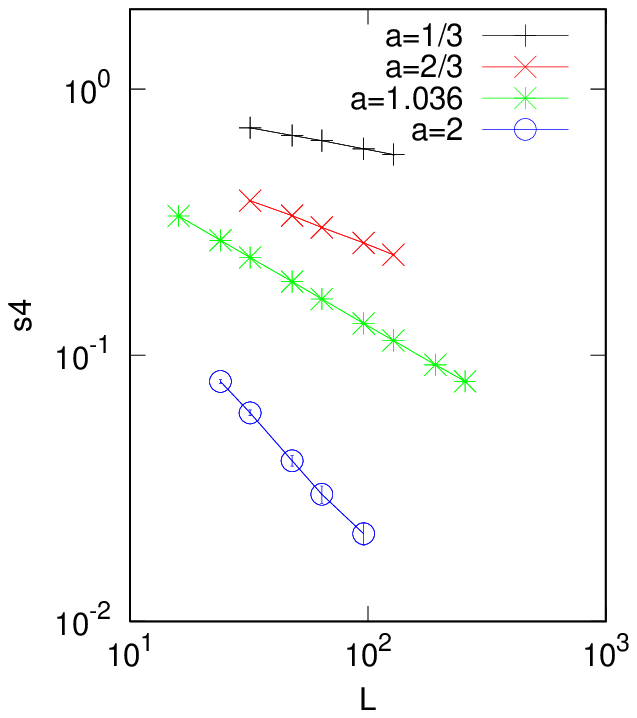}
\caption{Fluctuations $\Delta J$ (up to a factor $(J_1-J_2)/2$) of the
average random couplings versus the lattice size $L$. The different
curves correspond to different disorder correlations, i.e. to different
exponents $a=1/3$, $2/3$, $1.036$ and $2$ (from top to bottom).}
\label{Fig5}
\end{figure}

The variance of the average coupling $\Delta J_2$ is plotted on
figure~\ref{Fig5} versus the lattice size in the four cases $a=1/3$,
$2/3$, $1.036$ and $2$. Note that in the last two cases ($a\simeq 1.036$
and $a=2$), only the magnetisation in the two-dimensional section that
was used to construct the exchange couplings $J_{ij}$ is considered.
As expected, an algebraic decay with an exponent compatible with $a/2$
is observed. On figure~\ref{Fig2}, the collapse of the two peaks of the
magnetic susceptibility is observed for $a\simeq 1.036$ for lattice sizes
$L\sim {\cal O}(10^2)$ when $a\simeq 1.036$. According to figure~\ref{Fig5},
this corresponds to disorder fluctuations of order $\Delta J_2\simeq 0.06$.
For $a=1/3$ and $2/3$, none of the lattice sizes that were considered
correspond to so small disorder fluctuations. Indeed, $\Delta J_2=0.159(4)$
when $a=1/3$ for the largest lattice size $L=128$ and $\Delta J_2=0.090(3)$
when $a=2/3$. This strengthens the idea that the collapse will be
observed for larger lattice sizes for $a=1/3$ and $2/3$. Using the
scaling law (\ref{ScaleSigma}), one can even predict these sizes
to be of the order of $L^*\simeq 128(0.06/0.16)^{-2/a}\simeq 44,000$
for $a=1/3$ and $L^*\simeq 128(0.06/0.09)^{-2/a}\simeq 432$ for $a=2/3$.
On the other hand, disorder fluctuations are small for $a=2$ ($\Delta J_2
=0.0618(4)$ already for $L=24$), smaller than for $a\simeq 1.036$ with
lattice sizes $L\sim {\cal O}(10^2)$. 

These results do not depend on the quantity used to measure disorder
fluctuations. The scaling law (\ref{ScaleSigma}) suggests to use the
order parameter, polarisation $\overline{|\sigma_i\tau_i|}$
or magnetisation $\overline{|\sigma_i|}$, of the auxiliary models as an
alternative measure of the fluctuations of the couplings. This quantity will
be denoted $\Delta J_1$ in the following. $\Delta J_1$ and $\Delta J_2$ give
essentially the same information and, as can be seen on figure~\ref{Fig5},
take sensibly the same value, but $\Delta J_1$ presents the advantage of
being more stable numerically. More surprising is the fact that the same
conclusions can be drawn from the second contribution of disorder
to the partition function (\ref{DevPartF}). Expanding further, the next
term will involve the connected four-point correlation function
$\overline{J_iJ_jJ_kJ_l}_c$ of disorder. This quantity was assumed to
be irrelevant by Weinrib and Halperin. We considered the fourth-order
cumulant
   \begin{equation}
     \Delta J_4=\left(3\overline{\Big[{1\over N}\sum_{i,j} (J_{ij}-\bar J)
       \Big]^2}^2-\overline{\Big[{1\over N}\sum_{i,j} (J_{ij}-\bar J)
       \Big]^4}\right)^{1/4}.
   \end{equation}
As can be seen on figure~\ref{Fig5}, no qualitative difference between
the three quantities $\Delta J_1$, $\Delta J_2$ and $\Delta J_4$ is
observed.

% --- BINDER CUMULANT ---
%\begin{figure}
%\centering
%\psfrag{U}[Bc][Bc][1][1]{$U_M$}
%\psfrag{beta}[tc][tc][1][0]{$\beta$}
%\psfrag{L=24}[Bc][Bc][1][0]{\tiny $L=24$}
%\psfrag{L=32}[Bc][Bc][1][0]{\tiny $L=32$}
%\psfrag{L=64}[Bc][Bc][1][0]{\tiny $L=64$}
%\psfrag{L=48}[Bc][Bc][1][0]{\tiny $L=48$}
%\psfrag{L=96}[Bc][Bc][1][0]{\tiny $L=96$}
%\psfrag{L=128}[Bc][Bc][1][0]{\tiny $L=128$}
%\includegraphics[width=\FigSize]{Fig7a.eps}\quad
%\includegraphics[width=\FigSize]{Fig7b.eps}\par
%\includegraphics[width=\FigSize]{Fig7c.eps}\quad
%\includegraphics[width=\FigSize]{Fig7d.eps}
%\caption{Binder 4th-order cumulant of the 8-state Potts model
%with different disorder correlation exponents ($a=1/3$, $2/3$, $1.036$ and $2$
%from top to bottom and left to right). The different curves correspond to
%different lattice sizes.}
%\label{Fig7}
%\end{figure}

However, there are small differences between the four cases $a=1/3,2/3,1.036$
and $a$ that cannot be explained only in terms of disorder fluctuations.
The value of $\Delta J$ for the largest lattice size $L=128$ at $a=2/3$ is
close to the one estimated for $L=48$ at $a\simeq 1.036$. Therefore, the average
susceptibility should look qualitatively the same for $a=2/3$ at $L=128$
and for $a\simeq 1.036$ at $L=48$. It is not clear that it is indeed the
case on figure~\ref{Fig5}. Moreover, a nice collapse of the magnetic
susceptibility is observed at large $\beta$ for $a=1/3$ and $2/3$ while
it is not case for $a\simeq 1.036$ and 2. Stronger statements might be
formulated by comparing thermodynamic quantities displaying universal
properties. The natural candidate is the 4th-order Binder cumulant
     \begin{equation}
       U_M=1-{\overline{\langle m^4\rangle}
       \over 3\overline{\langle m^2\rangle^2}}
\end{equation}
whose value at the intersection of two curves with respect to
temperature is expected to be universal in the thermodynamic limit.
%The Binder cumulant is plotted on figure~\ref{Fig7}.
A notable difference between the different values of $a$ is that the
crossing points occur for inverse temperatures $\beta$ well below
$\beta_c=1$ when $a=1/3$ and $2/3$ and very close to $\beta_c=1$
when $a\simeq 1.036$ and $2$. Unfortunately, the error bars are large
and do not allow to be conclusive.

% --- RATIO RM ---
\begin{figure}[ht]
\centering
\psfrag{Rm}[Bc][Bc][1][1]{$\overline{\langle m\rangle^2}
/\overline{\langle m\rangle}^2-1$}
\psfrag{beta}[tc][tc][1][0]{$\beta$}
\psfrag{L=24}[Bc][Bc][1][0]{\tiny $L=24$}
\psfrag{L=32}[Bc][Bc][1][0]{\tiny $L=32$}
\psfrag{L=64}[Bc][Bc][1][0]{\tiny $L=64$}
\psfrag{L=48}[Bc][Bc][1][0]{\tiny $L=48$}
\psfrag{L=96}[Bc][Bc][1][0]{\tiny $L=96$}
\psfrag{L=128}[Bc][Bc][1][0]{\tiny $L=128$}
\includegraphics[width=\FigSize]{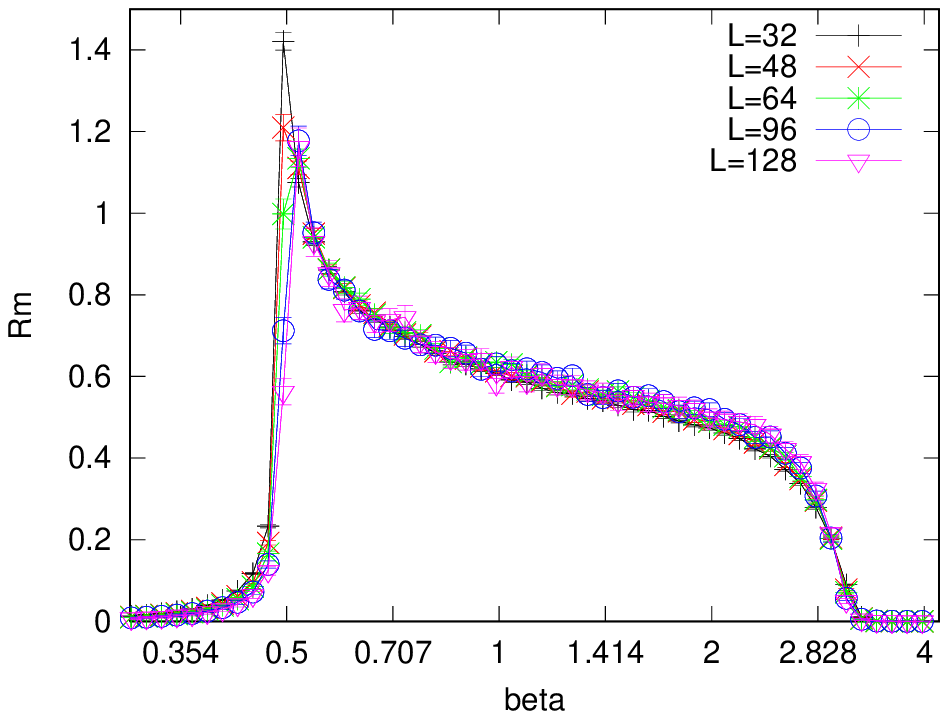}\quad
\includegraphics[width=\FigSize]{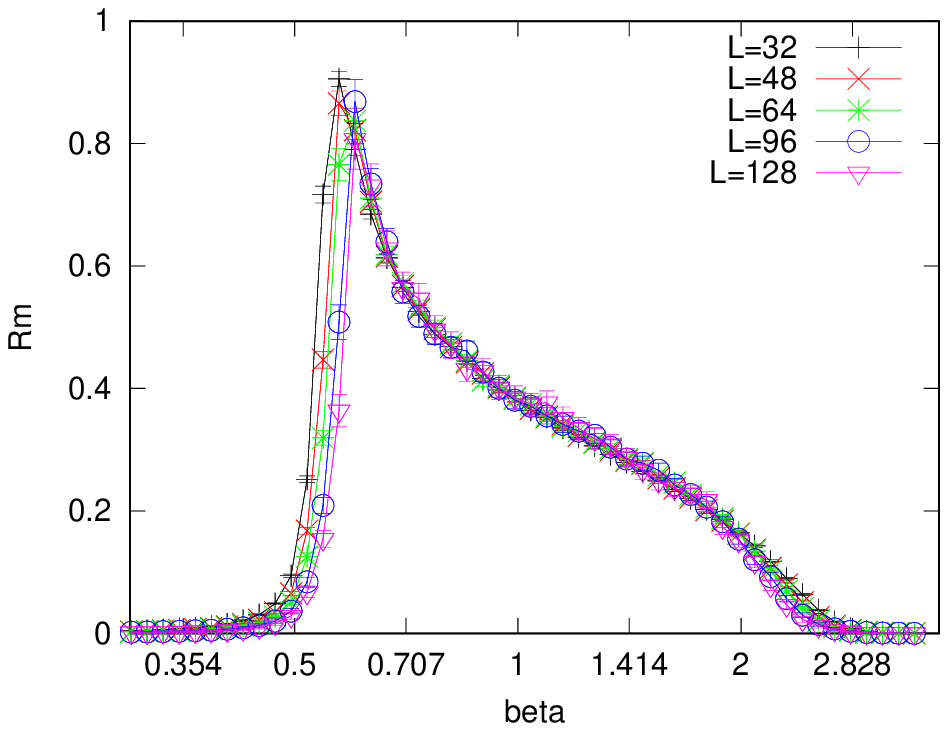}\par
\includegraphics[width=\FigSize]{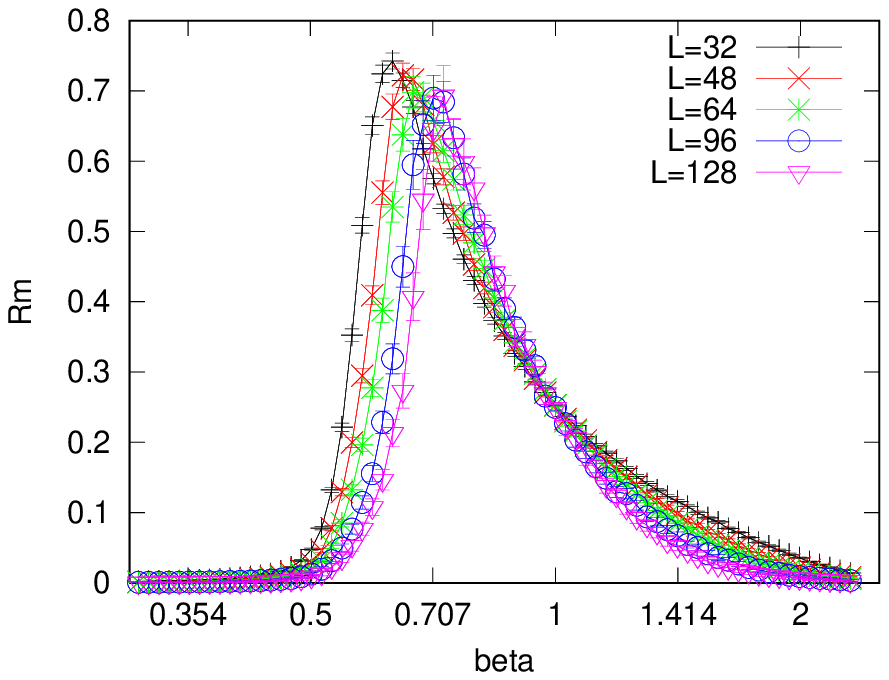}\quad
\includegraphics[width=\FigSize]{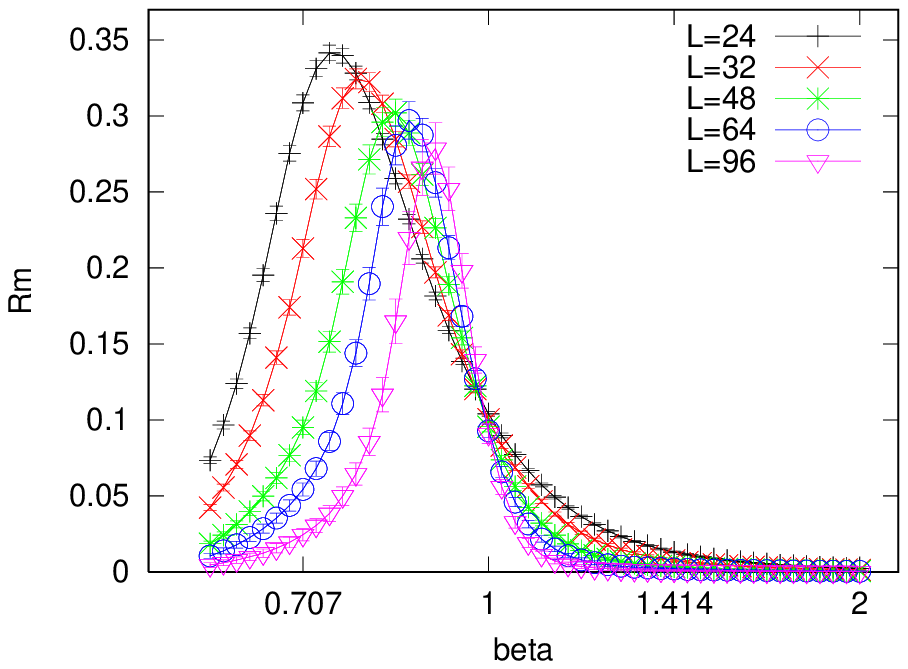}
\caption{Ratio $\overline{\langle m\rangle^2}/\overline{\langle m\rangle}^2-1$
of the 8-state Potts model with different disorder correlation exponents
($a=1/3$, $2/3$, $1.036$ and $2$ from left to right and top to bottom)
for a disorder strength $r=J_1/J_2=7.5$. The different curves correspond
to different lattice sizes.}\label{Fig3}
\end{figure}

Another quantity displaying universal properties is the ratio~\cite{Wiseman}
   \begin{equation}
     R_m={\overline{\langle m\rangle^2}-\overline{\langle m\rangle}^2
     \over \overline{\langle m\rangle}^2}
    \end{equation}
that measures the sample-to-sample fluctuations of magnetisation.
Outside of a critical point, all disorder realisations are expected
to lead to the same average magnetisation $\langle m\rangle$ in the
thermodynamic limit. Therefore, the ratio $R_m$ vanishes as $L\rightarrow
+\infty$ and magnetisation is said to be self-averaging. This is no longer
true at a fixed point where disorder is relevant. In this case, $R_m$
goes towards a finite value in the thermodynamic limit. This limit is
expected to be a universal quantity~\cite{Aharony}. Numerical data
for this ratio $R_m$ are plotted on figure~\ref{Fig3}. Two distinct
behaviours are observed. For $a=1/3$ and $a=2/3$, $R_m$ displays
a peak in the paramagnetic phase (small $\beta=1/k_BT$), followed by
a broad shouldering. The latter extends over a range of temperatures
which roughly corresponds to the range between the two peaks of the
average magnetic susceptibility (see figure~\ref{Fig2}). Interestingly,
the estimates of $R_m$ at any temperature in this shouldering are
compatible, within error bars, for all lattices sizes $L\in[32;128]$.
Unless a sudden decay of $R_m$ occurs at much larger lattices sizes,
we are led to the conclusion that magnetisation is a non-self-averaging
quantity in the whole range of temperatures between the two peaks
of the susceptibility. This conclusion is consistent with the assumption
of the existence of a Griffiths phase. On the other hand, for $a\simeq
1.036$ and $a=2$, the peak in the paramagnetic phase is softer and is
not followed by a shouldering but by a monotonous decay. More interesting
is the fact that the curves corresponding to different lattice
sizes cross each other at a single point, close to the self-dual point
$\beta_c=1$. This is consistent with the existence of a unique critical
point at $\beta_c=1$. Would it be possible that, in the case $a=2/3$,
the shouldering disappears at large lattice sizes to be replaced by
a monotonous decay with a single crossing point for different lattice
sizes? If the coupling fluctuations $\Delta J$ provides a measure of
the width of the Griffiths phase as discussed above, it should also
determine the range of temperatures around $\beta_c$ for which $R_m$
is finite and size-independent. Then the ratio $R_m$ should look
similar for $a=2/3$ at $L=128$ and for $a\simeq 1.036$ at $L=48$.
This is definitely not the case on figure~\ref{Fig3}. Therefore, the
Griffiths phase is not solely the consequence of disorder fluctuations
and there is no reason to expect the Griffiths phase to collapse into a
single point as $L^{-a/2}$.

\section{Griffiths phase and disorder strength}
All data presented in the previous section correspond to a disorder strength
$r=J_1/J_2=7.5$. Because the two peaks of the susceptibility were interpreted
as the ordering of macroscopic clusters with a majority of strong, or
weak, couplings, the disorder strength $r$ controls the width of the Griffiths
phase. One can therefore wonder whether disorder is not too strong
in the cases $a=1/3$ and $2/3$ which implies that a cross-over to the
Weinrib-Halperin fixed point would be observed at larger lattice sizes.
For the Potts model with uncorrelated disorder, strong scaling corrections
depending on $r=J_1/J_2$ were indeed observed. Accurate estimates of the critical
exponents became accessible only after an appropriate disorder strength was
determined. The by-far most
efficient technique was, in this case, to compute an effective central charge
$c_{\rm eff}$ by transfer matrix techniques and search for the maximum of
$c_{\rm eff}$. The central charge is unfortunately difficult to measure by
Monte Carlo simulations. Consequently, we will restrict ourselves to
observe the effect of a variation of the disorder strength $r$.

% --- FLUCTUATIONS DU COUPLAGE ---
\begin{figure}[ht]
\centering
\psfrag{s2}[Bc][Bc][1][1]{$(2-a)(\Delta J_2)^2 L^{a}$}
\psfrag{L}[tc][tc][1][0]{$L$}
\psfrag{a=1/3}[Bc][Bc][1][0]{\tiny $a=1/3$}
\psfrag{a=2/3}[Bc][Bc][1][0]{\tiny $a=2/3$}
\psfrag{a=1.036}[Bc][Bc][1][0]{\tiny $a\simeq 1.036$}
\psfrag{a=2}[Bc][Bc][1][0]{\tiny $a=2$}
\includegraphics[width=\FigSize]{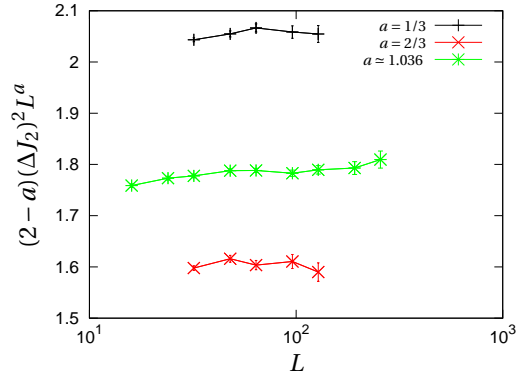}
\caption{Amplitude $w$ of disorder correlations (up to a factor $(J_1-J_2)^2/4$)
versus the lattice size $L$. The different curves correspond to different
disorder correlations, i.e. to different exponents $a=1/3$, $2/3$
and $1.036$.}
\label{Fig8}
\end{figure}

Weinrib and Halperin considered disorder correlations of the form
   \begin{equation}
     C(r_{ik})=\overline{J_{ij}J_{kl}}-\overline{J_{ij}}\ \!\overline{J_{kl}}
     =v\delta(\vec r_i-\vec r_k)+{w\over |r_{ik}|^a}
     \label{CorrelationWH}
   \end{equation}
where the two amplitudes $v$ and $w$ are irrelevant scaling fields at the
long-range random fixed point. When simulating a finite system with an
amplitude $w$ much larger (or much weaker) than the value $w^*$ taken at the
fixed point, the critical behaviour may be affected by strong scaling
corrections. Indeed, in the neighbourhood of the Weinrib-Halperin random
fixed point, the free energy density can be assumed to scale under
a dilatation with a scale factor $b$ as:
  \begin{equation}
    f(t,h,1/L,w)=b^{-d}f\big(b^{y_t}t,b^{y_h}h,b/L,b^{y_w}(w-w^*)\big)
  \end{equation}
where $t=|T-T_c|$ is the reduced temperature and $h$ the magnetic field.
At the critical point, i.e. $t=h=0$, and with $b=L$, the magnetic
susceptibility $\chi=-{\partial^2f\over\partial h^2}$ scales as
  \begin{equation}
    \chi(1/L,w)=L^{\gamma/\nu}{\cal F}\big(L^{y_w}(w-w^*)\big)
  \end{equation}
where $\gamma/\nu=2y_h-d$. The scaling function ${\cal F}$ involves a cross-over
length $\ell\sim (w-w^*)^{-1/y_w}$ associated to disorder. The dominant
finite-size scaling behaviour $\chi\sim L^{\gamma/\nu}$ will be hidden by
scaling corrections if $L\ll\ell$.

In the previous section, the fluctuations of the average coupling have been
compared for different exponents $a$. To compare now the amplitudes $w$ of
disorder correlations, note that integrating out disorder correlations leads
on one hand to
  \begin{equation}
    {1\over N^2}\sum_{ij,kl}\big[\overline{J_{ij}J_{kl}}
    -\overline{J_{ij}}\ \!\overline{J_{kl}}\big]
    \simeq {1\over L^2}\int_{L^2} C(\vec r)d^2\vec r
    \simeq {w\over L^2}\int_0^L {rdr\over r^a}
    =w{L^{-a}\over 2-a}
  \end{equation}
while on the other hand, the same quantity is equal to $(\Delta J_2)^2$
according to equation (\ref{DefSigma}).
%  \begin{equation}
%    {1\over N^2}\sum_{ij,kl}\big[\overline{J_{ij}J_{kl}}
%    -\overline{J_{ij}}\ \!\overline{J_{kl}}\big]
%    =\overline{\Big[{1\over N}\sum_{ij}(J_{ij}-\bar J)\Big]^2}
%    =(\Delta J_2)^2.
%  \end{equation}
The amplitude $w$ can therefore be recomputed as $w\simeq (2-a)
(\Delta J_2)^2L^a$. This estimate is plotted on figure~\ref{Fig8}.
Note that the amplitude $w$ is not plotted for $a=2$ because the definition
is inappropriate in this case (the integration of the correlations involves
a logarithm) and leads to $w=0$. As can be seen on figure~\ref{Fig8},
the amplitude $w$ does not evolve monotonously with $a$. This should not
be a surprise because the couplings have been generated from different
auxiliary models. The amplitude $w$ for $a\simeq 1.036$ lies in between the
amplitudes for $a=1/3$ and $a=2/3$. Therefore, the Griffiths phase
and the small $\nu$ exponents reported in~\cite{PRE} for $a=1/3$ and
$a=2/3$ cannot be explained as the result of strong scaling corrections.
Indeed, if one assume that the amplitude $w$ is close to $w^*$ in the case
$a\simeq 1.036$, which would explain why the collapse of the two peaks
of $\bar\chi$ is observed for reachable lattice sizes, one can conceive that
the Griffiths phase is the result of too strong disorder correlations,
i.e. $w>w^*$, in the case $a=1/3$. However, it is hard to understand how
weak correlations, i.e. $w<w^*$, would lead to a similar result
for $a=2/3$. The explanation in terms of a cross-over is therefore not
supported by the numerical data.

% --- SUSCEPTIBILITY AND RATIO AT Y=1.25 AND R=3.50 ---
\begin{figure}[ht]
\centering
\psfrag{chi}[Bc][Bc][1][1]{$\overline{\chi}$}
\psfrag{Rm}[Bc][Bc][1][1]{$\overline{\langle m\rangle^2}
/\overline{\langle m\rangle}^2-1$}
\psfrag{beta}[tc][tc][1][0]{$\beta$}
\psfrag{L=32}[Bc][Bc][1][0]{\tiny $L=32$}
\psfrag{L=64}[Bc][Bc][1][0]{\tiny $L=64$}
\psfrag{L=48}[Bc][Bc][1][0]{\tiny $L=48$}
\psfrag{L=96}[Bc][Bc][1][0]{\tiny $L=96$}
\psfrag{L=128}[Bc][Bc][1][0]{\tiny $L=128$}
\includegraphics[width=\FigSize]{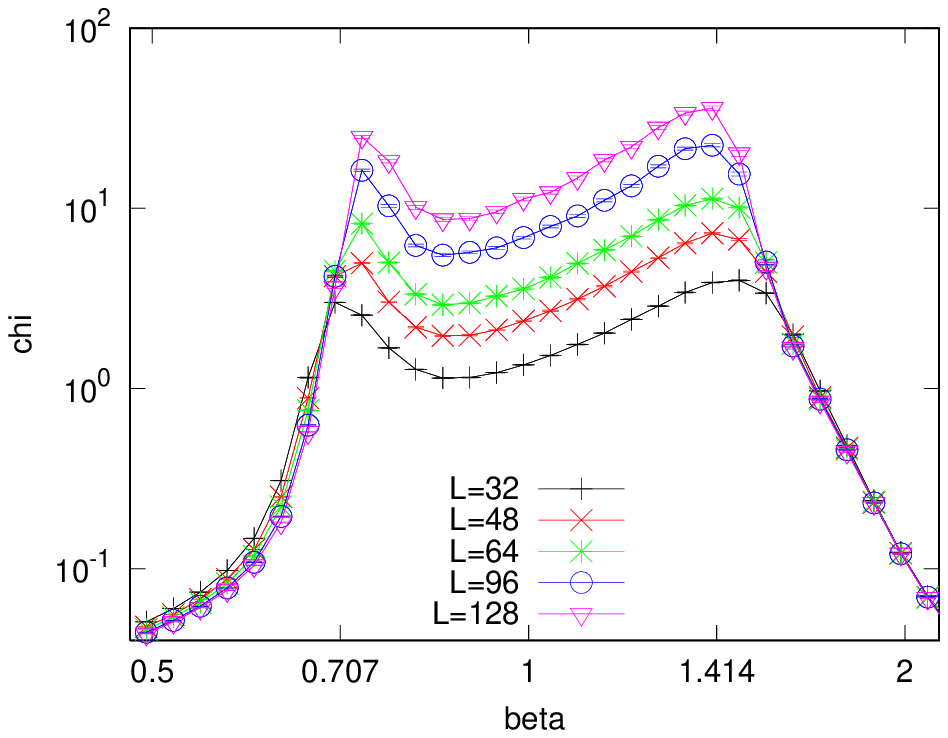}\quad
\includegraphics[width=\FigSize]{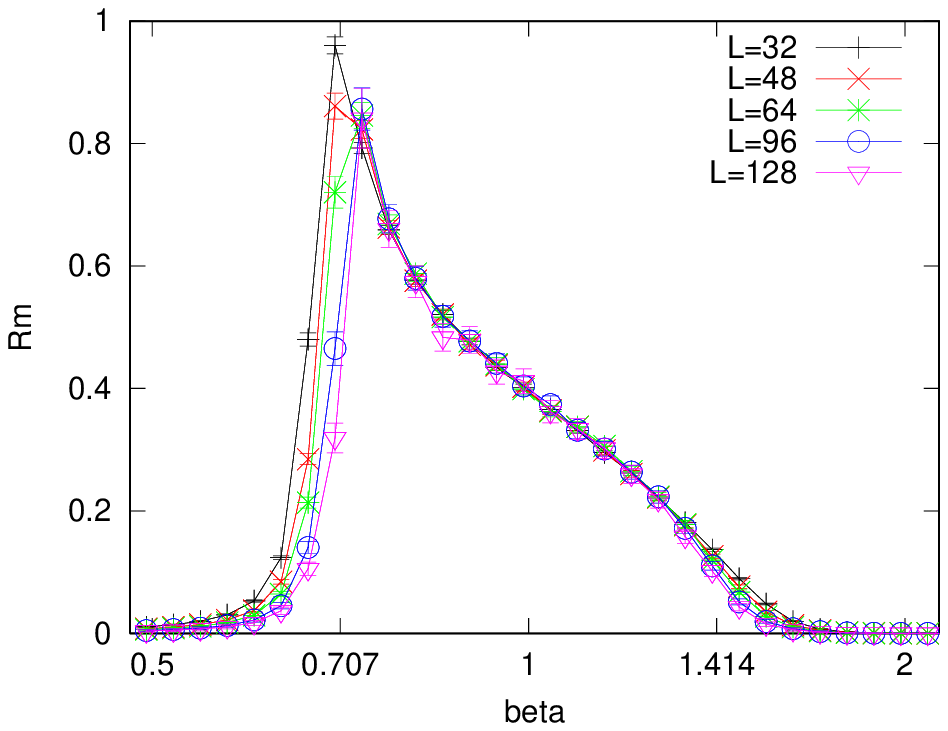}\par
\includegraphics[width=\FigSize]{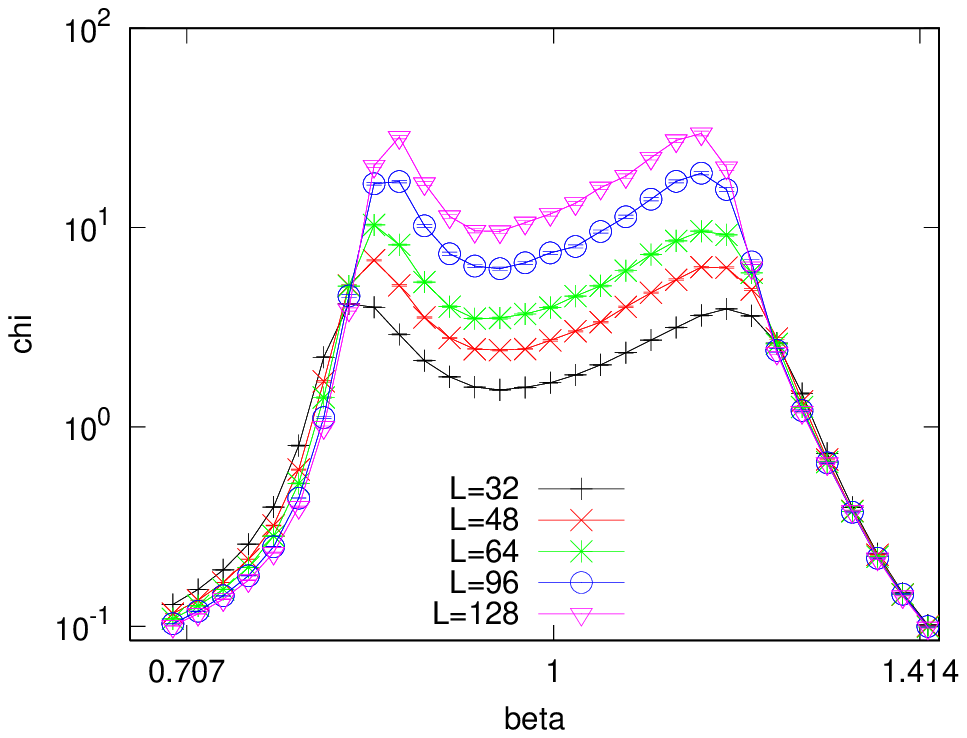}\quad
\includegraphics[width=\FigSize]{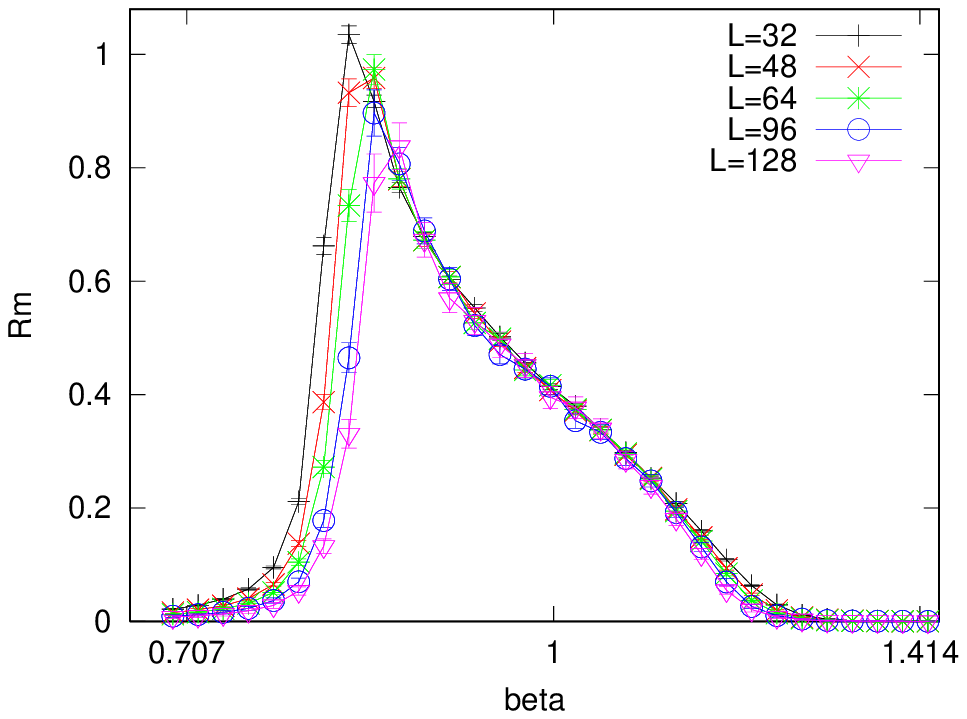}
\caption{On the left, average magnetic susceptibility of the 8-state Potts
model with a disorder correlation exponent $a=2/3$ and a disorder strengths
$r=J_1/J_2=3.5$ (top) and $r=2$ (bottom). The different curves correspond
to different lattice sizes. On the right, ratio $R_m=\overline{\langle m
\rangle^2}/\overline{\langle m\rangle}^2-1$ for the same models.
Note that the horizontal scales are not the same.}
\label{Fig9}
\end{figure}

However, as predicted by Weinrib and Halperin for the $\phi^4$ model, the
amplitude $w^*$ at the fixed point can be a function of $a$. In the following,
the effect of a change of the disorder strength, and therefore of $w-w^*$,
is studied in the case $a=2/3$. Since ${\cal F}$ depends on the scaling
variable $L^{y_w}(w-w^*)$, tuning the amplitude $w$ to come closer to $w^*$
is expected to be equivalent to increasing the lattice size $L$.
Therefore, the cross-over, if any, should be observed when $w$ is decreased.
On figure~\ref{Fig9}, the magnetic susceptibility is plotted for two
different disorder strengths $r=J_1/J_2=3.5$ and $r=2$. Comparing
with figure~\ref{Fig2} where the case $r=7.5$ was plotted, it is
clear that the width of the Griffiths phase is directly proportional
to the disorder strength $r$, and therefore the amplitude $w$. The
two peaks come closer as the disorder strength is reduced.
However, even for $r=2$, the magnetic susceptibility
is still very different from what is observed on figure~\ref{Fig2} for
$a\simeq 1.036$ and $r=7.5$. One can doubt that the two curves will look
similar for an even weaker disorder in the case $a=2/3$. Very probably,
the two peaks of the magnetic susceptibility for $a=2/3$ will collapse
but only when approaching $r=1$, i.e. the pure model.

The ratio $R_m=\overline{\langle m\rangle^2}/\overline{\langle m\rangle}^2-1$
provides a stronger evidence that what is observed for $a\simeq 1.036$
is not what should be expected for $a=2/3$ at weaker disorder.
As discussed when the figure~\ref{Fig2} was commented, the signature
of the Griffiths phase is a size-independent ratio $R_m$ over a finite
range of temperatures. In contrast, for $a\simeq 1.036$, the ratios $R_m$
computed at two different lattice sizes display a single crossing point at
a temperature evolving towards the self-dual point $\beta_c=1$. As observed
on figure~\ref{Fig9}, the diminution of the disorder strength induces
a reduction of the width of the Griffiths phase, and, as expected,
of the range of temperatures where the ratio $R_m$ appears to be
size-independent. However, the ratio $R_m$ looks surprisingly similar,
up to a temperature rescaling, at different disorder strengths.
Even at a disorder strength $r=2$, the behaviour is still very different
from what is observed for $a\simeq 1.036$.

% --- FLUCTUATIONS ---
%\begin{figure}
%\centering
%\psfrag{FluctM}[Bc][Bc][1][1]{$L^d[\overline{\langle m^2\rangle}
%-\overline{\langle m\rangle}^2]$}
%\psfrag{beta}[tc][tc][1][0]{$\beta$}
%\psfrag{L=24}[Bc][Bc][1][0]{\tiny $L=24$}
%\psfrag{L=32}[Bc][Bc][1][0]{\tiny $L=32$}
%\psfrag{L=64}[Bc][Bc][1][0]{\tiny $L=64$}
%\psfrag{L=48}[Bc][Bc][1][0]{\tiny $L=48$}
%\psfrag{L=96}[Bc][Bc][1][0]{\tiny $L=96$}
%\psfrag{L=128}[Bc][Bc][1][0]{\tiny $L=128$}
%\includegraphics[width=\FigSize]{Fig4a.eps}\quad
%\includegraphics[width=\FigSize]{Fig4b.eps}\par
%\includegraphics[width=\FigSize]{Fig4c.eps}\quad
%\includegraphics[width=\FigSize]{Fig4d.eps}
%\caption{Susceptibility $L^d[\overline{\langle m^2\rangle}
%-\overline{\langle m\rangle}^2]$ of the 8-state Potts model
%with different disorder correlation exponents ($a=1/3$, $2/3$, $1.036$ and $2$
%from top to bottom and left to right).
%The different curves correspond to different lattice sizes.}\label{Fig4}
%\end{figure}

\section{Conclusions}
New Monte Carlo simulations of the 2D 8-state Potts model with a
disorder involving algebraically decaying correlations $C(r)\sim r^{-a}$
with $a=1/3,2/3,1.036$ and $2$ are presented. While the analysis of the
magnetic susceptibility does not allow to exclude the possibility of a
collapse of the Griffiths phase into a single critical point, the study
of the self-averaging ratio $R_m=\overline{\langle m\rangle^2}/\overline{%
\langle m\rangle}^2-1$ allows to be more conclusive. Two different behaviours
are indeed observed for $a=1/3$ and $2/3$ on one hand and $a\simeq 1.036$
and $2$ on the other hand. The first case is compatible with the assumption
of the existence of a Griffiths phase while in the second case, the signature
of a single critical point is observed. These difference cannot be explained
by larger disorder fluctuations in the first case. Moreover, if the width of
the Griffiths phase depends on the disorder strength for $a=2/3$,
no cross-over towards the Weinrib-Halperin critical behaviour is observed
at weak disorder. These numerical results call for a theoretical understanding
of the precise mechanism behind this Griffiths phase. We hope that a theoretician
will find them sufficiently surprising to get interested into this problem.

\end{document}